\documentclass[aps,prb,twocolumn,groupaddress,showpacs]{revtex4}
\usepackage{subfigure, graphicx}
\begin{document}

\title{Environment-Mediated Quantum State Transfer}

\author{Yue Yin{*} and S. N. Evangelou}
\affiliation{Department of Physics, University of Ioannina, Ioannina 45 110, Greece}

\footnote{{*}on leave from National Laboratory of Solid State
Microstructures and Department of Physics, Nanjing University,
Nanjing 210093, China.}

\date{\today}

\begin{abstract}

We propose a scheme for quantum state transfer(QST) between two
qubits which is based on their individual interaction with a
common boson environment. The corresponding single mode spin-boson
Hamiltonian is solved by mapping it onto a wave propagation
problem in a semi-infinite ladder and the fidelity is obtained.
High fidelity occurs when the qubits are equally coupled to the
boson while the fidelity becomes smaller for nonsymmetric
couplings. The complete phase diagram for such an arbitrary QST
mediated by bosons is discussed.

\end{abstract}

\pacs {03.67.Hk,05.60.Gg,03.67.Mn}

\maketitle

\section {Introduction}

\par
\medskip
One of the most interesting problems in the area of quantum
information is how to transfer a quantum state from one location
to another. For example, QST from A to B can be done via quantum
teleportation\cite{r5} if a prior connection between the remote
places is established, by letting A and B have one from two
auxiliary entangled qubits. A more direct method should involve
flying qubits, such as photons in optical fibers, which can send
directly quantum information between two distant locations of a
quantum computer\cite{r6}. However, the latter approach would be
very difficult to realize experimentally since it requires an
interface between the optical system and the hardware where a
quantum computation takes place. A very successful QST was
pioneered in\cite{r1} which includes the locations A and B and
also the quantum transfer ground into the same system. In this
case the quantum information of the qubits at one end of the chain
propagate via the interaction between the components of a
permanently coupled physical system or a quantum graph. A perfect
or nearly perfect QST occurs between two local spins in quantum
spin chains and networks, at least for short
distances\cite{r1,r2,r3,r4}. The protocol for such quantum
communication relies on the fact that the auxiliary device which
plays the role of the quantum channel or quantum bus is the medium
itself. For example, the exchange interactions of a quantum spin
chain can allow transfer of quantum information between qubits
which belong to the first and last local spins while the rest of
the chain acts as the quantum channel. In this case the exchange
of information is achieved via magnon elementary excitations.
Physical devices for such QST could be built, at least in
principle, and since no external control is required they can
overcome possible decoherence mechanisms.

\par
\medskip
In this paper we show that a boson environment could be used to
transfer efficiently a quantum state by acting as a quantum
channel. It is known\cite{r7}  that entanglement can be introduced
between two qubits if both are independently coupled to a common
heat bath with many degrees of freedom. We  shall show that even
the simplest possible boson environment which consists of one mode
can also provide an efficient QST mechanism. For this purpose a
spin-boson Hamiltonian is introduced\cite{r8,r9} known for many
applications in physics and chemistry. A related spin-boson model
allowed Caldeira and Leggett \cite{r10} to study decoherence via
dissipation through a weak coupling of the spin to many bosons,
representing a universal realization of a physical environment.
Due to weak spin-boson interaction the excitations within the
boson heat bath could be ignored and the problem was solved,
leading to decoherence \cite{r10}. Our spin-boson model can be
regarded as an extension of \cite{r7} where two qubits coupled to
a common heat bath become entangled with each other. We show that
despite the absence of a direct interaction between them their
coupling to a simple boson environment mediates an efficient QST.
Environment mediated quantum control for a related multi-mode
system has been performed in \cite{r11}.

%\par
%\medskip
The proposed spin-boson model allows high fidelity QST between two
distant locations by choosing suitable parameters. In order to
make the problem tractable we chose the simplest possible quantum
channel which consists of a single-mode boson environment. This is
the first approximation to a full multi-mode Hamiltonian
considered in \cite{r8} by replacing the coupling to many modes by
a coupling to an effective boson. Our study proceeds as follows:
In chapter II the proposed spin-boson model is introduced with a
double two-level system Hamiltonian coupled to a single boson. In
chapter III  a formula is derived for the fidelity of a QST which
is obtained by mapping the system onto a wave-propagation problem
in a semi-infinite ladder. The results of our calculations with
the display of the corresponding phase-diagram and a discussion
about the efficiency of the scheme follow in chapter IV. Finally,
in chapter V we discuss possible extensions and applications.

\section {Model and Average Fidelity}

% Figure 1 : QST by quantum channel protocol
\begin{figure}
  \includegraphics[width=0.4\textwidth]{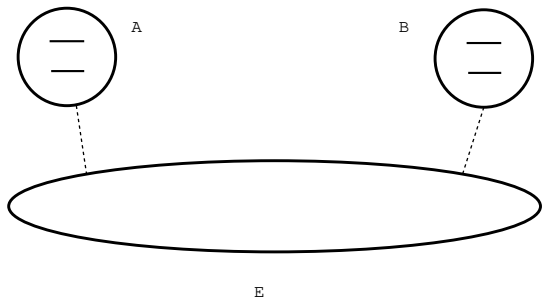}
  \caption{The proposed protocol for QST between two qubits A and B
   represented by two-level systems which interact with a common bosonic
   environment E acting as a quantum transfer channel. }
  \label{fig:system}
\end{figure}

\par
\medskip
The studied system is illustrated in Fig. \ref{fig:system}. The
qubits $A$ and $B$  are not directly coupled with each other but
are connected via an auxiliary boson environment $E$ both having
nonzero interaction with E. The qubits in A and B can be
represented by two local spins and E acts as the quantum channel.
Of course, if E is replaced by a quantum spin chain the model
reduces to that studied in \cite{r1}. The Hamiltonian is given by
the sum
\begin{eqnarray*}
  H & = & \omega^{0}_{A} \sigma^{z}_{A} + \omega^{0}_{B}\sigma^{z}_{B} + \omega b^{\dag}b \\
  & + & \lambda_{A} (b + b^{\dag})  \sigma^{x}_{A} + \lambda_{B} (b + b^{\dag})
  \sigma^{x}_{B}, \\
\end{eqnarray*}
with the qubits in A and B modeled by two-level systems of
separations $\omega^{0}_{A}$, $\omega^{0}_{A}$, the quantum
channel described by a single-boson mode environment of frequency
$\omega$ and nonzero linear couplings $\lambda_{A}$ and
$\lambda_{B}$ exist between the qubits and the boson channel E,
with $\sigma^{x/y/z}$ the corresponding Pauli matrices.  Note the
similarity of $H$ to a multi-mode model used to study entanglement
between the qubits in quantum control theory\cite{r11}. The main
differences between the present study and \cite{r7,r11} lies in
the number of modes and the presence or not of couplings between
the qubits and the quantum channel. We consider nonzero spin-boson
couplings $\lambda_{A}$ and $\lambda_{B}$ since they are expected
to be comparable to the two-level separations $\omega^{0}_{A}$ and
$\omega^{0}_{B}$.

\par
\medskip
The single-mode Hamiltonian $H$ although simple enough it cannot be
solved exactly. The Hilbert space consists of a direct product of
three parts with basis states
 $|\eta_{A}, \eta_{B}, m \rangle$, where, $\eta_{A/B}= 0, 1$ label the
qubits and $m=0, 1, 2, 3, ...$ is the single phonon excitation
number of the states in the quantum channel. The QST in this
system can be studied similarly to that in a spin
network\cite{r1}. Suppose that at time $t=0$ an unknown state $|
\psi_{A} \rangle = cos(\theta/2) | 0 \rangle + e^{i \phi}
sin(\theta/2) | 1 \rangle$ with parameters $\theta$, $\phi$, is
generated at qubit A and has to be transferred to B. We also
initialize the state of the qubit B to $| 0 \rangle$ and the state
of the quantum channel E to its lowest boson state $| 0 \rangle$.
The initial state of the whole system is $| \psi_{A}, 0, 0
\rangle$ which is separable. When evolution takes place the final
state at time $t$ in general becomes a non-separable mixed state.
The measurement of the state of qubit B is described by its
\emph{reduced density matrix} and both efficiency and quality of
the quantum communication is obtained by evaluation of the
corresponding \emph{fidelity}\cite{r1}.

\par
\medskip
The fidelity is usually computed  by taking average over all pure
input states $| \psi_{A} \rangle$ in its corresponding Bloch sphere
\begin{displaymath}
  \langle \mathcal{F}(t) \rangle = \frac{1}{4 \pi} \int d \Omega \langle
  \psi_{A} | \rho_{B}(t) | \psi_{A} \rangle,
\end{displaymath}
where the state of A to be transferred is $| \psi_{A}\rangle $,
$\rho_{B}(t)$ is the reduced density matrix of the qubit B at time
$t$ and the average is over all initial $|\psi_{A}\rangle$. If we
let the system evolve for a period of time $ t_{m}$, one can find
the maximum average fidelity $\langle \mathcal{F} \rangle_{m}$
from the time taken for the average fidelity to reach its first
peak corresponding to the maximum fidelity. The peak time $t_{p}$
is the second important quantity which can characterize a quantum
channel, the first being the average fidelity $\langle \mathcal{F}
\rangle_{m}$. High fidelity implies better quantum channel for QST
while shorter time to reach the peak means faster QST. If $\langle
\mathcal{F} \rangle_{m}$ becomes exactly unity we have perfect
QST\cite{c1} with the quantum state transferred from A to B
without any loss of quantum information.

\par
\medskip
The reduced density matrix for qubit B can be written
\begin{eqnarray*}
  \rho_{B} & = & Tr_{A,E}[\rho_{t}]
\end{eqnarray*}
by tracing out A and E of the evolved total density matrix
$\rho_{t} = U(t, 0) \rho_{0} U(0, t)$, with initial value $
\rho_{0} = \rho_{A} \otimes \rho_{B} \otimes \rho_{E}$ and time
evolution operator $U(t, 0) = e^{i H t}$, $\hbar=1$. This allows
to calculate the average fidelity for any time $t$, which we shall
simply call it fidelity from now on. As it stands this formula is
rather complicated to perform an analytic evaluation. In the next
chapter the problem is mapped onto an equivalent wave propagation
involving two ladders and the corresponding fidelity is written as
a function of waves propagating in these ladders.

\section { Wave Propagation}

\par
\begin{figure}
  \includegraphics[width=0.4\textwidth]{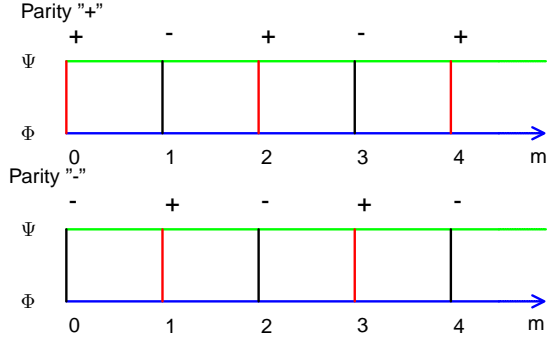}
  \caption{The equivalent wave propagation in  "+" and "-" ladders.
  The nodes denote the basis states and the lines the hoppings, the green lines
  denote $\sqrt{m}(\lambda_{A}+\lambda_{B})$, the blue lines
  $\sqrt{m}(\lambda_{A}-\lambda_{B})$, the red lines
  $\omega^{0}_{B} + \omega^{0}_{A}$, the black lines
  $\omega^{0}_{B} - \omega^{0}_{A}$ and the on-site energies
  are $\omega m$.}
  \label{fig:lattice}
\end{figure}

\medskip
A parity symmetry present in $H$ simplifies the Hamiltonian\cite{r9}
making it block-diagonal in a suitable two qubit Bell states basis
\begin{eqnarray*}
  | \Psi_{\pm}, m \rangle & = & \frac{1}{\sqrt{2}} ( | 0
  0 m \rangle \pm | 1 1 m \rangle ) \\
  | \Phi_{\pm}, m \rangle & = & \frac{1}{\sqrt{2}} ( | 0
  1 m \rangle \pm | 1 0 m \rangle ).\\
\end{eqnarray*}
The states split into two having  zero matrix elements between
each other and the block-diagonal Hamiltonian matrix is
illustrated via two decoupled ladders in Fig. \ref{fig:lattice}.
The states are represented by nodes and hoppings between the nodes
by the connecting lines. Note that the ladders of Fig. 2 are
rather similar to each other, their only difference being the
ordering of red and black lines. This becomes very helpful for our
calculation given in the Appendix where the computation is shown
to simplify in the chosen basis.

\par
\medskip
The obtained formula for the fidelity  can be given in the form
\begin{displaymath}
  \langle \mathcal{F} \rangle = \frac{1}{24} \sum_{m} ( Tr[A^{\dag}_{m} A_{m}]
  + Tr[B^{\dag}_{m} B_{m}] + Tr[C^{\dag}_{m} C_{m}] )
\end{displaymath}
with
\begin{eqnarray*}
  A_{m} & = & f_{m}(+,t) + \sigma^{z} f_{m}(-,t) \\
  B_{m} & = & f_{m}(+,t) + (-i \sigma^{y} ) f_{m}(-,t) \sigma^{x} \\
  C_{m} & = & \sigma^{z1} f_{m}(+,t) + \frac{\sigma^{+}}{2} f_{m}(+,t) \sigma^{x} \\
  & + & \frac{\sigma^{-}}{2} f_{m}(-,t) + \sigma^{z2} f_{m}(-,t) \sigma^{x} \\
  \sigma^{z1} & = & \left( \begin{array}{cc} 1 & 0 \\ 0 & 0 \\ \end{array} \right) \\
  \sigma^{z2} & = & \left( \begin{array}{cc} 0 & 0 \\ 0 & -1 \\ \end{array}
  \right),
\end{eqnarray*}
where $\sigma^{+,-,z}$ are the Pauli matrices and $f_{m}(\pm,t)$
is the propagator in the ladders shown. In the notation used, e.g.
$f_{3}(+,2)$ means the propagator from $m=0$ to slice $m=3$ at
time $t=2$ in the ladder with parity "+". This gives the fidelity
of QST written as a linear combination of the propagators in each
of the two ladders.

\par
\medskip
Since both ladders are semi-infinite the corresponding Hilbert
space must be truncated at a maximum phonon number $m$. In order
to approximate propagation for very long times long ladders with
large maximum $m$ are required. However, a careful study of the
 formula shows that the fidelity simply arises from the
\emph{difference} between propagators in the two ladders. For
example, for $\omega_{A}=0$ or $\omega_{B}=0$ the two ladders are
exactly the same and the fidelity becomes precisely zero. Since
their structure is rather similar, except for the ordering of
lines, if a wave reaches very far from the origin in one of them a
very small difference between the two propagators is expected with
no contribution to fidelity. Therefore, accurate computations of
fidelity do not require very long ladders and reasonable maximum
$m$ suffices, as seen in Fig. 3.

\begin{figure}
\subfigure[]{
\label{fig:chkm:sym}
\includegraphics[width=0.45\textwidth]{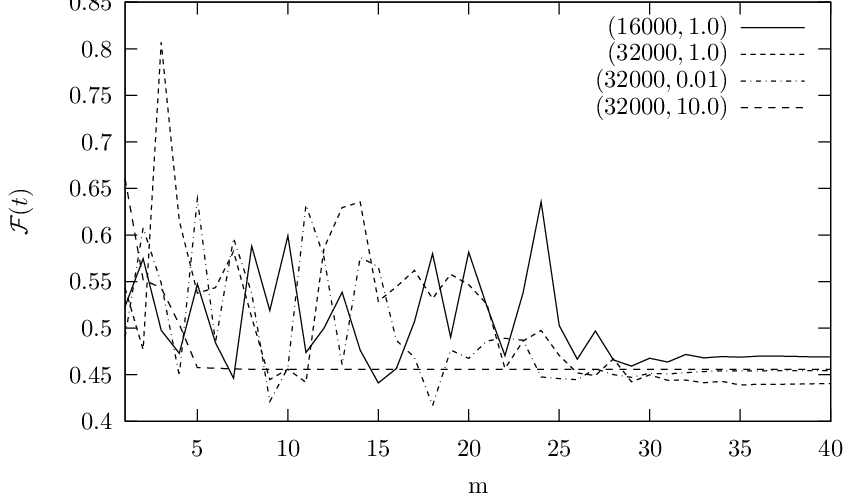}
}
\hspace{1in}
\subfigure[]{
\label{fig:chkm:nonsym}
\includegraphics[width=0.45\textwidth]{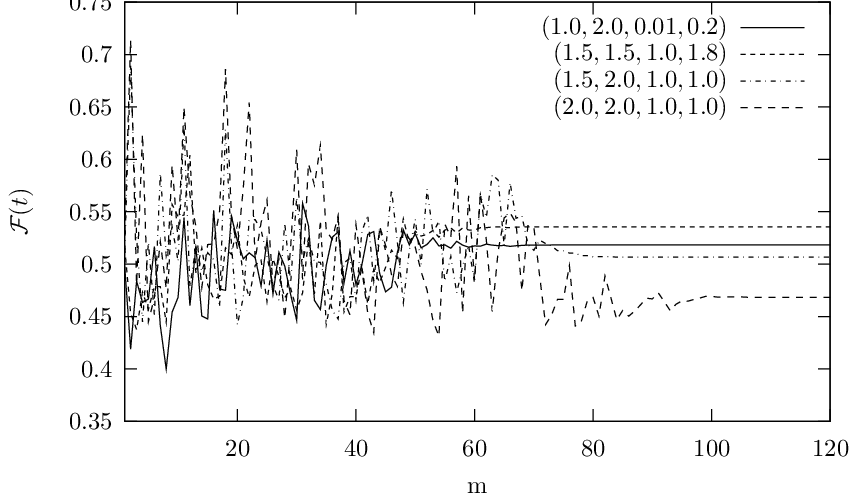}
}
  \caption{The convergence of the average fidelity $\mathcal{F}(t)$
   as a function of the maximum phonon number $m$: (a) symmetric case
   with $\lambda_{A}=\lambda_{B}=1.0$ and $\omega_{A}=\omega_{B}=\omega_{S}$ where
    the parameters in parentheses on the right of the figure are $(t, \omega_{S})$,
   and (b) non-symmetric case where the parameters displayed in
   parentheses are $(\lambda_{A}, \lambda_{B}, \omega_{A}, \omega_{B})$. Time $t$ is set to $32000$.}
   \label{fig:chkm}
\end{figure}

%\begin{figure}
%%\caption{Symmetric case}
%  \label{fig:chkm:sym} %% label for first subfigure
%  \includegraphics[width=0.45\textwidth]{chk_m_sym}
%  % convergence for symmetric case
%\end{figure}
%\begin{figure}
%  \label{fig:chkm:nonsym} %% label for first subfigure
%  \includegraphics[width=0.45\textwidth]{chk_m_non_sym}
%  % convergence for non-symmetric case
%  \caption{The convergence of the average fidelity $\mathcal{F}(t)$
%   as a function of the maximum phonon number $m$: (a) symmetric case
%   with $\lambda_{A}=\lambda_{B}=1.0$ and $\lambda_{A}=\lambda_{B}=1.0$ where
%    the parameters in parentheses on the right of the figure are $(t, \omega_{S})$,
%   and (b) non-symmetric case where the parameters displayed in
%   parentheses are $(t,\lambda_{A}, \lambda_{B}, \omega_{A}, \omega_{B})$.}
%%  \caption{Non-symmetric case. The convergence of the fidelity. $\mathcal{F}(t)$ is the average
%%   fidelity at time $t$. It is a function of the maximum phonon number $m$. (a)
%%    the symmetric case. $\lambda_{A}=\lambda_{B}=1.0$, and the parameters means
%%    $(t, \omega_{S})$. (b) the non-symmetric case. parameters means $(t,
%%    \lambda_{A}, \lambda_{B}, \omega_{A}, \omega_{B})$.}
%\end{figure}

\par
\medskip
The accuracy of the computed results is shown in Fig.
\ref{fig:chkm:sym} by plotting the fidelity as a function of the
maximum phonon number $m$ for the symmetric case with
$\lambda_{A}=\lambda_{B}=\lambda_{S}$,
$\omega_{A}=\omega_{B}=\omega_{S}$ and in Fig.
\ref{fig:chkm:nonsym} for the non-symmetric case. The fidelity is
shown to converge very rapidly for maximum phonon numbers $m=40$
or $50$ which permit to use reasonable coupling strengths. The
convergence does not depend on time $t_{m}$ and is also rather
insensitive to $\omega_{A/B}$ since it mostly depends on the
couplings $\lambda_{A/B}$. For example, the numerical results for
$\lambda_{S}=1.0$ and $\lambda_{S}=2.0$ required only $m=40$ to
$50$ and more that $m = 100$, respectively. In our computations
suitable maximum $m$ was chosen according to the values of
$\lambda_{A/B}$ and the convergence was checked by varying $m$.
For couplings $\lambda=0.0$ to $ 2$ and $\omega=0.0 $ to $80$ a
maximum phonon number $m$ between $50$ to $110$ was sufficient.

\section {Results and Discussion}

\begin{figure}
\subfigure[The phase diagram of the maximum fidelity
  as a function of equal couplings $\lambda_{S}$
  and equal qubit separations $\omega_{S}$. Three regions
  can be distinguished as explained in the text.]{
  \label{fig:fidsym:mfid}
  \includegraphics[angle=0, width=0.5\textwidth]{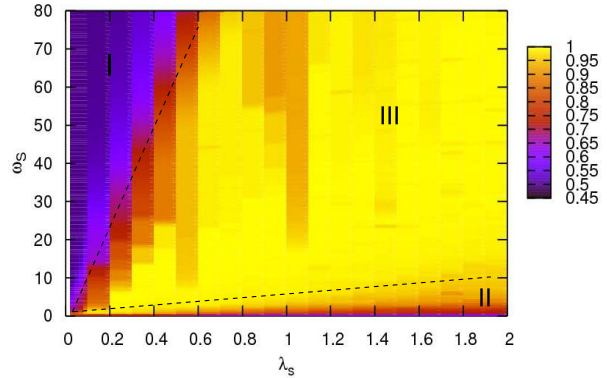}}
\hspace{1in}
\subfigure[The first peak time of the QST
  as a function of the equal couplings $\lambda_{S}$
  and the qubit separations $\omega_{S}$. This picture also
  has the three regions mentioned in the text.]
  { \label{fig:fidsym:pfm}
  \includegraphics[width=0.5\textwidth]{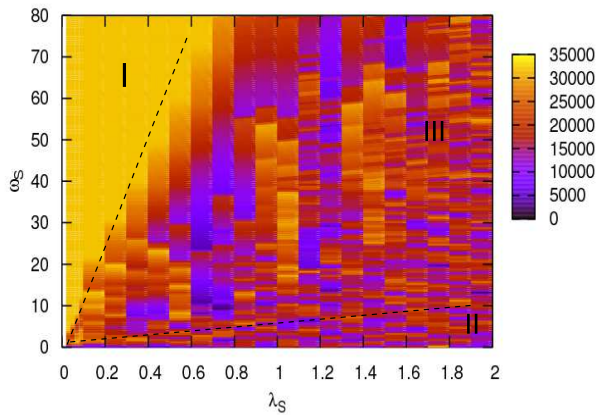}}
  \hspace{1in} \subfigure[The Region II of Fig.4(a) shown in
more detail.]
  {\label{fig:fidsym:mfidd} %% label for first subfigure
  \includegraphics[angle=0, width=0.45\textwidth]{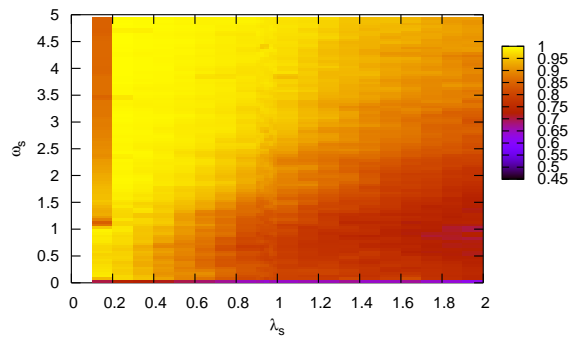}
}

\caption{Phase Diagram for the fidelity of QST} \label{fig:fidsym}
\end{figure}

\par
\medskip
The quality of QST is determined by the maximum of the average
fidelity $\langle \mathcal{F}\rangle_{m}$ in the time period from
$0$ to $ t_{m}$ and the time for occurrence of the first peak
$t_{p}$ when the system reaches its maximum. Higher fidelity means
more faithful state transfer while shorter peak time $t_{p}$
implies faster state transfer. The parameters $\omega_{A}$,
$\omega_{B}$, $\lambda_{A}$, $\lambda_{B}$ are taken in units of
$\omega=1$ while the maximum fidelity and the first peak time are
obtained in the time interval $[0, t_{m}=32\times 10^{3}]$.

\subsection {Phase Diagram for Symmetric Couplings}

\par
\medskip
The phase diagrams of the maximum fidelity and the first peak time
are shown  in Figs. 4(a), 4(b) as a function of the two parameters
$\omega_{S}=\omega_{A} = \omega_{B} $ and $\lambda_{S}=\lambda_{A}
= \lambda_{B}$. They can be divided into the following three
regions:

\par
\medskip
\textit{Region I}: a weak coupling region which lies in the upper
left corner of Fig. 4(a) where $\omega_{S}>>\lambda_{S}$. In this
case the corresponding first peak time $t_{p}$ shown in Fig. 4(b)
is large equal to the upper bound of the chosen time interval
$t_{m}$. In other words, the fidelity never reaches its maximum
within the adopted evolution time. This indicates that probably a
higher fidelity might occur for even longer times so that we can
call this a "slow region". We may conclude that a good state
transfer is impossible in this region because of the long times
$t_{p}$.

\par
\medskip
\textit{Region II}: lies in the lower part of the figure, which is
too small to be seen in Fig. 4(a) and this plot is magnified in
Fig. 4(c). In this region $\omega_{S}$ and $\lambda_{S}$ are of
the same order of magnitude so that the fidelity is again low but
for a different reason than that of region I. The first peak time
in this case from Fig. 4(b) is less than $t_{m}$ and the QST is
affected by increasing $\omega_{S}$. For zero $\omega_{S}$  no QST
is possible while it becomes better when increasing  the
qubit-environment coupling $\lambda_{S}$.

\par
\medskip
\textit{Region III}: The rest of Fig. 4(a). One can see that in
the majority of this region high fidelity occurs with the first
peak time mostly being  less than $5\times 10^{3}$. This region
corresponds to a two-valley Hamiltonian and the system behaves as
a good quantum channel.

\subsection {Phase diagram for non-Symmetric Couplings}

\begin{figure}
\subfigure[]{
% \caption{$\delta\omega$}
  \label{fig:fidnonsym:dw} %% label for first subfigure
  \includegraphics[width=0.5\textwidth]{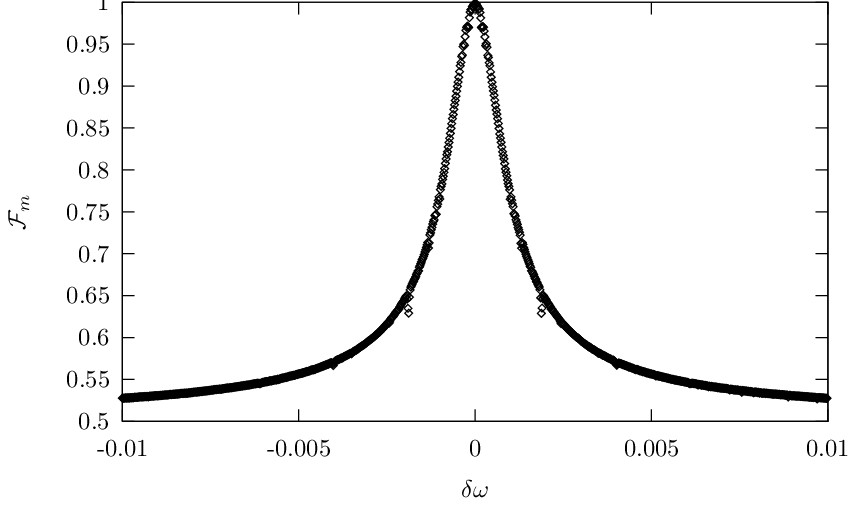}
% convergence for symmetric case
}
\hspace{1in}
\subfigure[]{
  \label{fig:fidnonsym:dl} %% label for first subfigure
  \includegraphics[width=0.5\textwidth]{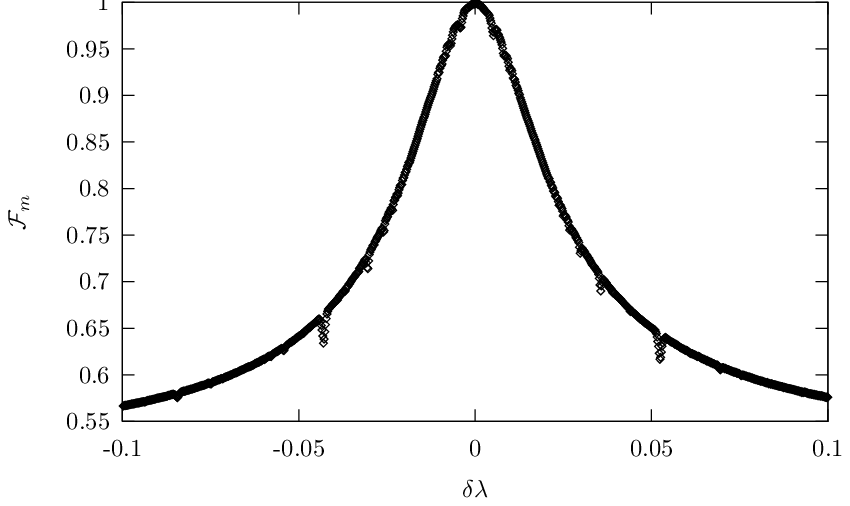}
  % convergence for non-symmetric case
}
  \caption{The maximum of the average fidelity $\langle
  \mathcal{F}\rangle_{m}=0.998$,
   a value close to a perfect QST, becomes lower
    for deviations from equal frequencies  $\omega_{A}=\omega_{B}=20.0$ and
    equal couplings $\lambda_{A}=\lambda_{B}=0.8$,
   with $ \delta\omega= \omega_{B} - \omega_{A}$  and
    $ \delta\lambda = \lambda_{B} - \lambda_{A}$,
     respectively. The $\langle \mathcal{F}\rangle_{m}$ is obtained
    in the region $[0, 33\times 10^{3}]$.
    }
  \label{fig:fignonsym}
\end{figure}

\par
\medskip
We have also considered the non-symmetric case where the two
couplings and the two frequencies are not equal. The influence of
a deviation from equal couplings is studied by choosing
$\lambda_{S}=0.8, \omega_{S}=20.0$ with the corresponding point
 of the symmetric phase diagram belonging to region III having very
high fidelity equal to $0.998$. A small deviation $\delta\omega$
in $\omega_{B}$ with $ \omega_{B} = \omega_{A} + \delta\omega$ is
shown in Fig. 5(a) to influence dramatically the QST,  which is
extremely sensitive even for deviations of the order of $10^{-4}$.
The asymmetry in the coupling constants is shown in Fig. 5(b) to
have a much smaller effect.

\section {Conclusions}
\par
\medskip
Although the role of an environment is usually that of causing
decoherence for a quantum system the presence of entanglement
between the system and the environment also signals the
possibility that quantum information can be transferred via the
environment. We suggest a QST between two qubits via a coupling to
a common boson medium which acts as the quantum channel. We have
derived a formula for the corresponding fidelity of the state
transfer by mapping this problem into a wave propagation, which is
much easier to understand and solve. For symmetric couplings and
frequency separation case high fidelity QST between the two qubits
is obtained for a wide range of parameters. We show that small
deviations from this symmetry can dramatically lower the QST.

\par
\medskip
Questions for further study are: (i) possible extensions of the
present scheme to include a multimode boson environment since our
results can cover only approximately the multimode case, (ii)
connections of QST to wave propagation in media also in the
presence of disorder which can also give ballistic, chaotic and
even localized states (in the latter case QST is impossible) and
(iii) possible realization of an experiment where QST mediated by
bosons can occur, for example, between two quantum dots coupled to
the appropriate phonon environment of a nanostructure.

\section {Acknowledgment}

This work was supported by Marie Curie RTN NANO No 504574
"Fundamentals in Nanostructures".

\section {Appendix: Derivation of the Formula of Fidelity}

The average fidelity
\begin{displaymath}
  \langle \mathcal{F}(t) \rangle = \frac{1}{4 \pi} \int d \Omega \langle \psi_{A} | \rho_{B}(t) | \psi_{A} \rangle
\end{displaymath}
over $| \psi_{A} \rangle$ becomes
\begin{eqnarray*}
  | \psi_{A} \rangle & = & cos(\frac{\theta}{2}) | 0 \rangle + e^{i
    \phi} sin(\frac{\theta}{2}) | 1 \rangle \\
  \frac{1}{4 \pi} \int d \Omega .. & = & \frac{1}{4 \pi} \int^{\pi}_{0} d
  sin(\theta) d \theta \int^{2 \pi}_{0} d \phi ... \\
\end{eqnarray*}
The reduced density matrix $\rho_{B}(t)$ can be calculated via
\begin{eqnarray*}
  \rho_{B} & = & Tr_{A,E}[\rho_{t}] \\
  \rho_{t} & = & U(t, 0) \rho_{0} U(0, t) \\
  U(t, 0) & = & e^{i H t} \\
\end{eqnarray*}
where, the partial trace over the degrees of freedom for qubit A and
the quantum channel E is taken. $H$ is the Hamiltonian for the
system $A \otimes B \otimes E$ and $U(t, 0)$ is the corresponding
time evolution operator.

To simplify the formula first we have calculated the integral. It is
convenient for us to choose \emph{coherent vector
  representation} \cite{r6} to express the density matrix.
\begin{eqnarray*}
  \rho_{B}(t) & = & \frac{1}{2} (1 + \vec{p}_{B}(t) \cdot \vec{\sigma}) \\
  \rho_{A} & = & \frac{1}{2} (1 + \vec{p}_{A} \cdot \vec{\sigma}) \\
\end{eqnarray*}
an assuming the relation between two coherent vectors
\begin{displaymath}
  \vec{p}_{B}(t) = T(t) \cdot \vec{p}_{A} + \vec{T}_{0}(t)
\end{displaymath}
we can carry out the integral
\begin{displaymath}
  \langle \mathcal{F}(t) \rangle = \frac{1}{2} [1 + \frac{1}{3} Tr(T(t))
  ].
\end{displaymath}

We need to calculate the matrix $T(t)$, e.g., to express the final
state of qubit B as a function of initial state of qubit A
\begin{displaymath}
  \rho_{B}(t) = Tr_{A,E}[U(t, 0) \rho(0) U(0, t)]
\end{displaymath}
where $\rho(0)$ is the initial state of the whole system ($A \otimes
B \otimes E$), it is separable so that
\begin{eqnarray*}
  \rho(0) & = & \rho_{A}(0) \otimes \rho_{B}(0) \otimes \rho_{E}(0) \\
  \rho_{B}(0) & = & | 0 \rangle \langle 0 | \\
  \rho_{E}(0) & = & | 0 \rangle \langle 0 |. \\
\end{eqnarray*}
By inserting $|\eta_{A}, \eta_{B}, m \rangle$ into these formulae we
find
\begin{displaymath}
  \rho_{B}(\eta_{B}, \eta^{'}_{B}, t) =
  \sum_{\bar{\eta}_{A}, \bar{\eta^{'}}_{A}} J_{B}(\eta_{B}, \eta^{'}_{B},
  \bar{\eta}_{A}, \bar{\eta^{'}}_{A}, t) \rho_{A}(\bar{\eta}_{A},
  \bar{\eta^{'}}_{A}, 0)
\end{displaymath}
\begin{eqnarray*}
  J_{B}(\eta_{B}, \eta^{'}_{B}, \bar{\eta}_{A}, \bar{\eta^{'}}_{A}, t) & = & \\
  \sum_{\eta_{A}} J(\eta_{A}\eta_{B}, \eta_{A}\eta^{'}_{B}, t; \bar{\eta}_{A} 0,
  \bar{\eta^{'}}_{A} 0, 0) & & \\
\end{eqnarray*}
\begin{eqnarray*}
  J(\eta_{A}\eta_{B}, \eta^{'}_{A}\eta^{'}_{B}, t; \bar{\eta}_{A} \bar{\eta}_{B},
  \bar{\eta^{'}}_{A} \bar{\eta^{'}}_{B}, 0) & = & \\
  \sum_{m} \langle \bar{\eta}_{A} \bar{\eta}_{B}, 0 | U(0, t) | \eta_{A}
  \eta_{B}, 0 \rangle \langle \eta^{'}_{A} \eta^{'}_{B}, 0 | U(t, 0) |
  \bar{\eta^{'}}_{A} \bar{\eta^{'}}_{B}, 0 \rangle, & & \\
\end{eqnarray*}
where $\eta_{A/B}=0/1$, $m = 0, 1, 2, 3, ...$

\par
\medskip
The matrix element between $\rho_{B}(t)$ and $\rho_{A}(0)$ is
related by the function $J_{B}$
\begin{displaymath}
  T = \left(
    \begin{array}{ccc}
      T^{x}(01)+T^{x}(10) & i [T^{x}(10)-T^{x}(01)] & T^{x}(00)-T^{x}(11) \\
      T^{y}(01)+T^{y}(10) & i [T^{y}(10)-T^{y}(01)] & T^{y}(00)-T^{y}(11) \\
      T^{z}(01)+T^{z}(10) & i [T^{z}(10)-T^{z}(01)] & T^{z}(00)-T^{z}(11)
    \end{array}
  \right)
\end{displaymath}
where,
\begin{eqnarray*}
  T^{x}(\eta \eta^{'}) & = & J_{B}(01, \eta \eta^{'}) + J_{B}(10, \eta \eta^{'}) \\
  T^{y}(\eta \eta^{'}) & = & i [J_{B}(10, \eta \eta^{'}) - J_{B}(01, \eta \eta^{'}) ] \\
  T^{z}(\eta \eta^{'}) & = & J_{B}(00, \eta \eta^{'}) - J_{B}(11, \eta \eta^{'})
\end{eqnarray*}
By going into Bell basis the final expression for the fidelity is
obtained.

%% section 9


\begin{thebibliography}{13}
\bibitem{c1} An average fidelity equal to unity means the final state
  is different from the initial state to be transfered by a phase
  factor. Since this phase is known via a simple rotation the final state
  becomes exactly the same as the initial state.
\end{thebibliography}
\end{document}